\documentclass[pra,aps,showpacs,twocolumn,longbibliography]{revtex4}
\UseRawInputEncoding

\usepackage{amssymb}
\usepackage{amsmath}
\usepackage{mathtools}
\usepackage[dvipdfm]{hyperref}
\usepackage{hyperref}
\usepackage{stmaryrd}
\usepackage{float}
\usepackage{caption}

\allowdisplaybreaks
\begin{document}

\title{Kirkwood-Dirac classical pure states}
\author{Jianwei Xu}
\email{xxujianwei@nwafu.edu.cn}
\affiliation{College of Science, Northwest A$\&$F University, Yangling, Shaanxi 712100,
China}

\begin{abstract}
Kirkwood-Dirac (KD) distribution is a representation of quantum states.
Recently, KD distribution has been found applications in many areas such as
in quantum metrology, quantum chaos and foundations of quantum theory. KD
distribution is a quasiprobability distribution, and negative or nonreal
elements may signify quantum advantages in certain tasks. A quantum state is
called KD classical if its KD distribution is a probability distribution.
Since most quantum information processings use pure states as ideal resources, then a key
problem is to determine whether a quantum pure state is KD classical. In
this paper, we provide some characterizations for the general structure of
KD classical pure states. As an application of our results, we prove a
conjecture raised by De Bi\`{e}vre [Phys. Rev. Lett. 127, 190404 (2021)]
which finds out all KD classical pure states for discrete Fourier
transformation.
\end{abstract}

\maketitle

\section{Introduction}

Kirkwood-Dirac (KD) distribution is a representation of quantum states,
which was discovered by Kirkwood in 1933 \cite{PR-1-33-Kirkwood}, and by
Dirac in 1945 \cite{RMP-1945-Dirac}. In this paper, we only consider the KD
distribution of pure states, since in most quantum processing tasks we use
pure states as ideal resources.\ Suppose $A=\{|a_{j}\rangle \}_{j=1}^{d}$
and $B=\{|b_{k}\rangle \}_{k=1}^{d}$ are two orthonormal bases of a $d$%
-dimensional complex Hilbert space $\mathbb{C}^{d}.$ For a pure state $%
|\psi \rangle $ in $\mathbb{C}^{d},$ the KD distribution of $|\psi \rangle $ with
respect to $A$ and $B$ is defined as
\begin{equation}
Q_{jk}(|\psi \rangle )=\langle a_{j}|\psi \rangle \langle \psi |b_{k}\rangle \langle
b_{k}|a_{j}\rangle.  \label{eq1.1}
\end{equation}%
Under this definition, when $|\psi \rangle $ is normalized, we certainly have $\sum_{j=1}^{d}Q_{jk}(|\psi \rangle )=|\langle \psi |b_{k}\rangle |^{2},$ $\sum_{k=1}^{d}Q_{jk}(|\psi \rangle )=|\langle
a_{j}|\psi \rangle |^{2}$ and $\sum_{j,k=1}^{d}Q_{jk}(|\psi \rangle )=1.$ $\{|\langle
\psi |b_{k}\rangle |^{2}\}_{k=1}^{d}$ and $\{|\langle a_{j}|\psi \rangle
|^{2}\}_{j=1}^{d}$ are all probability distributions, we then call $%
\{Q_{jk}(|\psi \rangle )\}_{j,k=1}^{d}$ a quasiprobability distribution since $%
Q_{jk}(|\psi \rangle )$ may be negative or nonreal numbers. We call $|\psi \rangle $
KD classical if $Q_{jk}(|\psi \rangle )\geq 0$ for all $j,k.$ Otherwise, we call $%
|\psi \rangle $ KD nonclassical. Remark that, $Q_{jk}(|\psi \rangle )$ just coincides with the Bargmann invariant of three pure states \cite{JMP-1964-Bargmann,PRA-1999-Rabei,JDM-2016-Chien}. Ref. \cite{arXiv-2021-Oszmaniec} reviewed some recent progress about Bargmann invariant.

In recent years, KD distribution has been found significant applications in several
areas of quantum mechanics, such as in weak-value amplification \cite%
{32,33,34,35,36,37,38,39,40,41,42,43,arxiv-2023-Wagner,JPA-2023-Budiyono}, tomographic reconstruction of a quantum state \cite{44,45,46,47,48}, quantum chaos \cite{49,50,51,52,53,54}, quantum metrology \cite{55,56,57,PRL-2022-NYH,PRA-2023-Das}, quantum thermodynamics \cite{51,59,60,Quantum-2022-Lostaglio}, quantum coherence \cite{PRA-2023-Budiyono,arxiv-2023-Budiyono}, and the foundations of quantum mechanics \cite{12,41, 61,62,63,64,65,66,67,68,69,PLA-2021-Ban,JPA-2023-Budiyono-2,arXiv-2023-Rastegin}, see
\cite{JPA-2021-NYH} for an extensive review and references therein. In these
scenarios, negative or nonreal elements of KD distribution may signify
quantum advantages. Since most quantum information processings use pure states as ideal resources, therefore it becomes a key problem to determine the
general structure of KD classical pure states.

To address this issue, we define the transition matrix $U^{AB}$ and support
uncertainty $n_{A}(\psi )+n_{B}(\psi )$. The transition matrix $U^{AB}$ of $%
A $ and $B$ is defined by its entries $U^{AB}_{jk}$ as
\begin{equation}
U_{jk}^{AB}=\langle a_{j}|b_{k}\rangle .  \label{eq1.2}
\end{equation}%
$U^{AB}$ is unitary and $\langle b_{k}|a_{j}\rangle =(U_{jk}^{AB})^{\ast }$
appear in Eq. (\ref{eq1.1}), here $($ $)^{\ast }$ stands for the complex
conjugate of $($ $)$. We also use $($ $)^{t}$ to stand for the transpose of $%
($ $);$ use $\cdot ,$ $||()||$ to denote the standard inner product and
standard norm in $\mathbb{C}^{d};$ and use $|($ $)|$ to denote the absolute value of $%
($ $)$ in the set of complex numbers $\mathbb{C}$ or the numbers of elements in the set $%
().$ The support uncertainty $n_{A}(\psi )+n_{B}(\psi )$ is defined by the $%
A $ support of $|\psi \rangle ,$ $n_{A}(\psi ),$ and $B$ support of $|\psi \rangle ,
$ $n_{B}(\psi ),$ as
\begin{eqnarray}
n_{A}(\psi ) &=&|\{j\in \llbracket{1,d}\rrbracket \Big|\langle a_{j}|\psi \rangle
\neq 0\}|,  \label{eq1.3} \\
n_{B}(\psi ) &=&|\{k\in  \llbracket{1,d}\rrbracket \Big|\langle b_{k}|\psi \rangle
\neq 0\}|,  \label{eq1.4}
\end{eqnarray}
where $\llbracket{j_{1},j_{2}}\rrbracket$ represents the set of consecutive integers from $%
j_{1}$ to $j_{2}.$ For example, $\llbracket{2,5}\rrbracket=\{2,3,4,5\}.$

In Ref. \cite{JPA-2021-NYH}, the authors proved that when pure state $|\psi
\rangle $ is KD classical and $|\psi \rangle \langle \psi |\notin
\{|a_{j}\rangle \langle a_{j}|\}_{j=1}^{d},$ $|\psi \rangle \langle \psi
|\notin \{|b_{k}\rangle \langle b_{k}|\}_{k=1}^{d},$ then
\begin{equation}
n_{A}(\psi )+n_{B}(\psi )\leq \frac{3}{2}d.  \label{eq1.5}
\end{equation}%
De Bi\`{e}vre showed that \cite{PRL-2021-Bievre} when all elements of $%
U^{AB}=\{\langle a_{j}|b_{k}\rangle \}_{j,k=1}^{d}$ are nonzero and $|\psi
\rangle $ is KD classical, then
\begin{equation}
n_{A}(\psi )+n_{B}(\psi )\leq d+1.  \label{eq1.6}
\end{equation}%
De Bi\`{e}vre also conjectured that \cite{PRL-2021-Bievre} when $%
U^{AB}=\{\langle a_{j}|b_{k}\rangle \}_{j,k=1}^{d}$ form a discrete Fourier
transformation (DFT) $\{\langle a_{j}|b_{k}\rangle =\frac{1}{\sqrt{d}}e^{i%
\frac{2\pi }{d}jk}\}_{j,k=1}^{d}$ with $i=\sqrt{-1}$, then $|\psi \rangle $
is KD classical iff (if and only if)
\begin{equation}
n_{A}(\psi )n_{B}(\psi )=d.  \label{eq1.7}
\end{equation}
Some relations between the transition matrix and support uncertainty were discussed in Ref. \cite{PRA-2022-Xu}.

In this paper we provide some characterizations for the general structure of KD classical pure
states. The rest of this paper is organized as follows. In Sec. II, we establish general
structure for KD classical pure states. In Sec. III, we explore the relation
between the support uncertainty and the number of zeros in transition
matrix. In Sec. IV, we give some examples to demonstrate the applications of our results. Sec. V is a brief summary.

\section{General structure of KD classical pure states}

In this section, we establish two structure theorems for KD classical pure
states. We state Facts 1 to 4, they are immediate consequences
of the definition $Q_{jk}(|\psi \rangle )$ but useful for subsequent
discussions.

\textbf{Fact 1.} All $\{Q_{jk}(|\psi \rangle )\}_{j,k=1}^{d}$ keep invariant if
we replace $A=\{|a_{j}\rangle \}_{j=1}^{d}$ by $\widetilde{A}=\{e^{i\xi
_{j}}|a_{j}\rangle \}_{j=1}^{d}$ and replace $B=\{|b_{k}\rangle \}_{k=1}^{d}$
by $\widetilde{B}=\{e^{i\eta _{k}}|b_{k}\rangle \}_{k=1}^{d},$ where $\{\xi
_{j}\}_{j=1}^{d}\subseteq \mathbb{R},$ $\{\eta _{k}\}_{j=1}^{d}\subseteq \mathbb{R},$ $\mathbb{R}$ is
the set of all real numbers.

\textbf{Fact 2.}
\begin{equation}
Q_{jk}(e^{i\theta }|\psi \rangle )=Q_{jk}(|\psi \rangle )\text{ for any }%
\theta \in \mathbb{R}.
\label{eq2.1}
\end{equation}

\textbf{Fact 3.} When $n_{A}=1,$ the KD classical states are $%
\{e^{i\alpha _{j}}|a_{j}\rangle \}_{j=1}^{d}$ with $\{\alpha
_{j}\}_{j=1}^{d}\subseteq \mathbb{R}.$ For the KD classical state $e^{i\alpha
_{j}}|a_{j}\rangle ,$ we have $n_{B}=|\{k \Big|\langle b_{k}|a_{j}\rangle \neq
0\}|.$ The case of $n_{B}=1$ is similar.

\textbf{Fact 4.} Suppose $A=A_{1}\cup A_{2},$ $B=B_{1}\cup
B_{2},$ $A_{1}\neq \varnothing ,$ $A_{2}\neq \varnothing ,$ $B_{1}\neq
\varnothing ,$ $B_{2}\neq \varnothing ,$ span$A_{1}=$span$B_{1},$ span$A_{2}=$span$B_{2}.$  Then the pure state $|\psi \rangle $ is
KD classical with respect to $A$ and $B$, iff
\begin{equation}
|\psi \rangle =|\psi _{1}\rangle +|\psi _{2}\rangle ,  \label{eq2.2}
\end{equation}%
with $|\psi _{1}\rangle \in$span$A_{1}$  being KD classical with respect to $A_{1}$ and $%
B_{1},$ $|\psi _{2}\rangle \in$span$A_{2}$  being KD classical with respect to $A_{2}$ and $%
B_{2}.$ Here span$A_{1}$ denotes the linear span of $A_{1}$ over the set of complex numbers.

Fact 1 implies that $|\psi \rangle $ is KD classical with respect to $%
\{|a_{j}\rangle \}_{j=1}^{d}$ and $\{|b_{k}\rangle \}_{k=1}^{d},$ iff $|\psi
\rangle $ is KD classical with respect to $\{e^{i\xi _{j}}|a_{j}\rangle
\}_{j=1}^{d}$ and $\{e^{i\eta _{k}}|b_{k}\rangle \}_{k=1}^{d}.$ Fact
2 implies that $|\psi \rangle $ is KD classical iff $e^{i\theta }|\psi
\rangle $ is KD classical.

Fact 3 determines the structure of the KD classical states with $%
n_{A}(\psi )=1$ (or $n_{B}(\psi )=1$), we call such states $%
\{e^{i\alpha _{j}}|a_{j}\rangle \}_{j=1}^{d}$ the basis states of $A$, the basis states of $B$ are similar. Thus below we mainly consider the
case of $\min \{n_{A},n_{B}\}\geq 2.$ Fact 3 explicitly shows that for fixed positive integers $(n_{A},n_{B})$, there may not exist a pure state $|\psi\rangle $ such that $n_{A}(\psi )=n_{A}$ and $n_{B}(\psi )=n_{B}.$ For example, for DFT of dimension $d$, there does not exist a pure state $|\psi\rangle $ such that $n_{A}(\psi )=1$ and $n_{B}(\psi )<d.$

Under Fact 4, to determine the KD classical pure states of $A$ and $B$, we
first consider whether the transition matrix $U^{AB}$ can be decomposed as
the direct sum of $U^{AB}=U^{A_{1}B_{1}}\oplus U^{A_{2}B_{2}}.$ If so, we
only need to determine the KD classical pure states of $U^{A_{1}B_{1}}$ and $%
U^{A_{2}B_{2}},$ and then use Eq. (\ref{eq2.2}). Below we focus on the case that the transition matrix $U^{AB}$ can not be decomposed into the form of direct sum.

Theorem 1 below shows the general structure of KD classical pure states. To
state Theorem 1, we write $\langle a_{j}|\psi \rangle ,$ $\langle \psi
|b_{k}\rangle $ and $U_{jk}^{AB}$ in the polar form as
\begin{eqnarray}
\langle a_{j}|\psi \rangle  &=&|\langle a_{j}|\psi \rangle |e^{i\alpha _{j}}%
\text{ if }\langle a_{j}|\psi \rangle \neq 0, \\
\langle \psi |b_{k}\rangle  &=&|\langle \psi |b_{k}\rangle |e^{i\beta _{k}}%
\text{ if }\langle \psi |b_{k}\rangle \neq 0, \\
U_{jk}^{AB} &=&|U_{jk}^{AB}|e^{i\theta _{jk}}\text{ if }U_{jk}^{AB}\neq 0,
\end{eqnarray}
where $\alpha _{j},\beta _{k},\theta _{jk}\in \mathbb{R}.$

\textbf{Theorem 1.} A pure state $|\psi \rangle $ is KD classical with respect
to $A$ and $B$, iff there exist nonempty sets $S_{A}\subseteq
\llbracket{1,d}\rrbracket,$ $S_{B}\subseteq \llbracket{1,d}\rrbracket,$ $%
\{\alpha _{j}\}_{j\in S_{A}}\subseteq \mathbb{R},$ $\{\beta _{k}\}_{k\in
S_{B}}\subseteq \mathbb{R},$ $\{A_{j}>0\}_{j\in S_{A}},$ $\{B_{k}>0\}_{k\in S_{B}},$
such that
\begin{eqnarray}
\theta _{jk} &\equiv &\alpha _{j}+\beta _{k}\text{ mod }2\pi \text{ when }%
Q_{jk}(|\psi \rangle )\neq 0,  \label{eq2.3} \\
|\psi \rangle &=&\sum_{j\in S_{A}}A_{j}e^{i\alpha _{j}}|a_{j}\rangle
=\sum_{k\in S_{B}}B_{k}e^{-i\beta _{k}}|b_{k}\rangle.  \label{eq2.4}
\end{eqnarray}%

\emph{Proof.} Expanding $|\psi \rangle $ in the bases $A$ and $B,$ evidently
there exist $\varnothing \neq S_{A}\subseteq \llbracket{1,d}\rrbracket,$ $%
\varnothing \neq S_{B}\subseteq \llbracket{1,d}\rrbracket,$ $\{A_{j}>0\}_{j\in
S_{A}},$ $\{B_{k}>0\}_{k\in S_{B}}$ such that Eq. (\ref{eq2.3},\ref{eq2.4}) hold.

Obviously $Q_{jk}(|\psi \rangle )=\langle a_{j}|\psi \rangle \langle \psi |b_{k}\rangle
\langle b_{k}|a_{j}\rangle =0$ when $j\in \llbracket{1,d}\rrbracket\backslash
S_{A}$ or $k\in \llbracket{1,d}\rrbracket \backslash S_{B}.$ For $j\in S_{A}$ and
$k\in S_{B},$ if $\langle a_{j}|b_{k}\rangle =0$ then $Q_{jk}(|\psi \rangle )=0;$ if $%
\langle a_{j}|b_{k}\rangle \neq 0$ and $Q_{jk}(|\psi \rangle )>0,$ then $e^{i\alpha
_{j}}e^{i\beta _{k}}e^{-i\theta _{jk}}=1$ and $\theta _{jk}\equiv \alpha
_{j}+\beta _{k} \text{ mod }2\pi .$ The claim then follows.
$\hfill\blacksquare$

The freedom of global phase for pure KD classical state $|\psi \rangle $ in
Fact 1 corresponds to the invariance of Eq. (\ref{eq2.3})\ under the
transformation: $\alpha _{j}\rightarrow \alpha _{j}+\theta $ for all $j\in
S_{A},$ $\beta _{k}\rightarrow \beta _{k}-\theta $ for all $k\in S_{B},$
with $\theta \in \mathbb{R}.$ Note that Eq. (\ref{eq2.4}) implies
\begin{eqnarray}
A_{j} &=&\sum_{k\in S_{B}}B_{k}|\langle a_{j}|b_{k}\rangle |,j\in S_{A};
\label{eq2.5} \\
B_{k} &=&\sum_{j\in S_{A}}A_{j}|\langle a_{j}|b_{k}\rangle |,k\in S_{B}.
\label{eq2.6}
\end{eqnarray}%
Further, when $A$ and $B$ are mutually unbiased bases (MUBs), i.e., $%
\{|\langle a_{j}|b_{k}\rangle |=1/\sqrt{d}\}_{j,k=1}^{d},$ Eqs. (\ref{eq2.5},\ref{eq2.6}) imply Corollary 1 below.

\textbf{Corollary 1.} Suppose $A$ and $B$ are MUBs, then the pure state $|\psi \rangle $ is KD
classical with respect to $A$ and $B$ iff
\begin{eqnarray}
&&n_{A}(\psi )n_{B}(\psi )=d.  \label{eq2.9}
\end{eqnarray}

\emph{Proof.} In \cite{PRL-2021-Bievre}, De Bi\`{e}vre showed that when
\begin{equation}
n_{A}(\psi )n_{B}(\psi )=\frac{1}{M_{AB}^{2}},  \label{eq4.3}
\end{equation}
then $|\psi \rangle $ is KD classical, where $M_{AB}=\max \{|\langle
a_{j}|b_{k}\rangle |\}_{j,k=1}^{d}.$ For MUBs, Eq. (\ref{eq4.3}) becomes $%
n_{A}(\psi )n_{B}(\psi )=d.$ Conversely, for MUBs, if $|\psi \rangle $ is KD
classical, then Eqs. (\ref{eq2.5},\ref{eq2.6}) imply Eq. (\ref{eq2.9}).
$\hfill\blacksquare$

Note that for MUBs in prime dimension, Fact 3 and Corollary 1 together imply that the only pure KD classical states are the basis states.

We further analyze Eqs. (\ref{eq2.5},\ref{eq2.6}). Write $V=\{|\langle a_{j}|b_{k}\rangle |\}_{j\in S_{A},k\in S_{B}}$ in the
form of matrix, $\overrightarrow{A}=(A_{1},A_{2},...,A_{n_{A}})^{t},$ $%
\overrightarrow{B}=(B_{1},B_{2},...,B_{n_{B}})^{t},$ then Eqs. (\ref{eq2.5},\ref{eq2.6}) imply that
\begin{eqnarray}
\overrightarrow{A} &=&V\overrightarrow{B},  \label{eq2.10} \\
\overrightarrow{B} &=&V^{t}\overrightarrow{A},  \label{eq2.11} \\
\overrightarrow{A} &=&VV^{t}\overrightarrow{A},  \label{eq2.12} \\
\overrightarrow{B} &=&V^{t}V\overrightarrow{B}.  \label{eq2.13}
\end{eqnarray}
Eqs. (\ref{eq2.10},\ref{eq2.11},\ref{eq2.12},\ref{eq2.13}) will be used in the proof of Theorem 2.

Note that Fact 3 is a special case of Theorem 1. In
Theorem 1, when $\min \{n_{A},n_{B}\}\geq 2,$ Eq. (\ref{eq2.3}) implies that
\begin{equation}
\theta _{j_{1}k}-\theta _{j_{2}k}\equiv \alpha _{j_{1}}-\alpha _{j_{2}}\text{
mod }2\pi   \label{eq2.14}
\end{equation}
when $\langle a_{j_{1}}|b_{k}\rangle \langle a_{j_{2}}|b_{k}\rangle \neq 0,$
$\forall j_{1},j_{2}\in S_{A},$ $k\in S_{B};$
\begin{equation}
\theta _{jk_{1}}-\theta _{jk_{2}}\equiv \beta _{k_{1}}-\beta _{k_{2}}\text{
mod }2\pi   \label{eq2.15}
\end{equation}%
when $\langle a_{j}|b_{k_{1}}\rangle \langle a_{j}|b
_{k_{2}}\rangle \neq 0,$ $j\in S_{A},$ $\forall k_{1},k_{2} \in S_{B}.$

Eq. (\ref{eq2.14}) means that $\theta _{j_{1}k}-\theta _{j_{2}k}$ is
independent of $k,$ that is
\begin{equation}
\theta _{j_{1}k_{1}}-\theta _{j_{2}k_{1}}\equiv \theta _{j_{1}k_{2}}-\theta
_{j_{2}k_{2}}\text{ mod }2\pi   \label{eq2.16}
\end{equation}%
when $\langle a_{j_{1}}|b_{k_{1}}\rangle \langle a_{j_{1}}|b_{k_{2}}\rangle
\langle a_{j_{2}}|b_{k_{1}}\rangle \langle a_{j_{2}}|b_{k_{2}}\rangle \neq 0,
$ $\forall j_{1},j_{2} \in  S_{A},$ $\forall k_{1},k_{2} \in  S_{B}.$ Eq. (%
\ref{eq2.15}) means that $\theta _{jk_{1}}-\theta _{jk_{2}}$ is independent
of $j,$ that is
\begin{equation}
\theta _{j_{1}k_{1}}-\theta _{j_{1}k_{2}}=\theta _{j_{2}k_{1}}-\theta
_{j_{2}k_{2}}\text{ mod }2\pi   \label{eq2.17}
\end{equation}%
when $\langle a_{j_{1}}|b_{k_{1}}\rangle \langle a_{j_{1}}|b_{k_{2}}\rangle
\langle a_{j_{2}}|b_{k_{1}}\rangle \langle a_{j_{2}}|b_{k_{2}}\rangle \neq 0,
$ $\forall j_{1},j_{2} \in  S_{A},$ $\forall k_{1},k_{2} \in  S_{B}.$
Also, Eq. (\ref{eq2.16}) is obviously equivalent to Eq. (\ref{eq2.17}), then
Eq. (\ref{eq2.14}) is equivalent to Eq. (\ref{eq2.15}). Therefore Eq. (\ref%
{eq2.16}) or Eq. (\ref{eq2.17}) provides a necessary and sufficient condition to check Eq. (\ref%
{eq2.3}).

Theorem 2 below characterizes the
structure of KD classical pure states via the transition matrix. Suppose $|\psi \rangle $ is a KD classical
state, then $|\psi \rangle $ is of the form in Eq. (\ref{eq2.4}). We relabel
$A=\{|a_{j}\rangle \}_{j=1}^{d}$ and $B=\{|b_{k}\rangle \}_{k=1}^{d}$ such
that $S_{A}=\{j\}_{j=1}^{n_{A}}$ and $S_{B}=\{k\}_{k=1}^{n_{B}}.$ We can
further reorder $S_{A}=\{j\}_{j=1}^{n_{A}}$ and $S_{B}=\{k\}_{k=1}^{n_{B}},$
such that the submatrix $\binom{1,2,...,n_{A};}{1,2,...,n_{B}.}$ is block
diagonal and exhibits maximum number of nonzero blocks, here
\begin{equation}
\binom{1,2,...,n_{A};}{1,2,...,n_{B}.} \nonumber
\end{equation}
denotes the
submatrix formed by the $\{j\}_{j=1}^{n_{A}}$ rows and $\{k\}_{k=1}^{n_{B}}$
columns of $U^{AB}.$ Then using Fact 1, we can choose $\widetilde{A}%
=\{e^{i\alpha _{j}}|a_{j}\rangle \}_{j=1}^{n_{A}}\cup \{|a_{j}\rangle
\}_{j=n_{A}+1}^{d}$ and $\widetilde{B}=\{e^{-i\beta _{k}}|b_{k}\rangle
\}_{k=1}^{n_{B}}\cup \{|b_{k}\rangle \}_{k=n_{B}+1}^{d}$ such that $\langle
\widetilde{a_{j}}|\psi \rangle \geq 0,$ $\langle \psi |\widetilde{b_{k}}%
\rangle \geq 0$ with $|\widetilde{a_{j}}\rangle =e^{-i\alpha
_{j}}|a_{j}\rangle ,$ $|\widetilde{b_{k}}\rangle =e^{i\beta
_{k}}|b_{k}\rangle $ for all $S_{A}=\{j\}_{j=1}^{n_{A}}$ and $%
S_{B}=\{k\}_{k=1}^{n_{B}}.$ $\{Q_{jk}(|\psi\rangle)\geq 0\}_{j=1,k=1}^{n_{A},n_{B}}$ then
results in $\{\langle \widetilde{a_{j}}|\widetilde{b_{k}}\rangle \geq
0\}_{j=1,k=1}^{n_{A},n_{B}}$ and $\binom{1,2,...,n_{A};}{1,2,...,n_{B}.}%
=\oplus _{j=1}^{s}R_{\geq 0}^{(j)}$ with each $R_{\geq 0}^{(j)}$ having nonnegative elements. We depict such $U^{AB}$ in Fig. 1. Since the column
(row) vectors of $U^{AB}$ are orthogonal to each other, then any two column
(row) vectors in distinct matrices of $\{C_{j}\}_{j=1}^{s}$ ($%
\{R_{j}\}_{j=1}^{s}$) are orthogonal. Furthermore we have Theorem 2 below.

\textbf{Theorem 2.}  Suppose $|\psi \rangle $ is a KD classical
state, and $U^{AB}$ is as in Fig. 1. Then  $\{R_{\geq 0}^{(j)}\}_{j=1}^{s},$ $%
\{C_{j}\}_{j=1}^{s},$ $\{R_{j}\}_{j=1}^{s},$ $n_{A}(\psi )$ and $n_{B}(\psi
) $ in Fig. 1 have the relations
\begin{eqnarray}
\text{rank}(C_{j}) &=&\text{cols}(R_{\geq 0}^{(j)})-1,  \label{eq2.29} \\
\text{rank}(R_{j}) &=&\text{rows}(R_{\geq 0}^{(j)})-1,  \label{eq2.30} \\
n_{A}(\psi )+n_{B}(\psi ) &\leq &d+s,  \label{eq2.31} \\
s &\leq &\frac{d}{2}.  \label{eq2.32}
\end{eqnarray}%
where rows$(R_{\geq 0}^{(j)})$ (cols$(R_{\geq 0}^{(j)})$) stands for the
number of rows (columns) of matrix $R_{\geq 0}^{(j)}.$ When rows$(R_{\geq
0}^{(j)})=1$ then $R_{j}=0,$ when cols$(R_{\geq 0}^{(j)})=1$ then $C_{j}=0.$
When  cols$(R_{\geq 0}^{(j)})\geq 2$ (rows$(R_{\geq 0}^{(j)})\geq 2$),
then $C_{j}$ ($R_{j}$) has no zero column (row) vector, also the column
(row)  vectors of $C_{j}$ ($R_{j}$) can not be divided into two nonempty
sets with orthogonal spanned spaces.

\begin{figure}[!ht]
\includegraphics[width=0.45\textwidth,bb=180 80 800 520,clip]{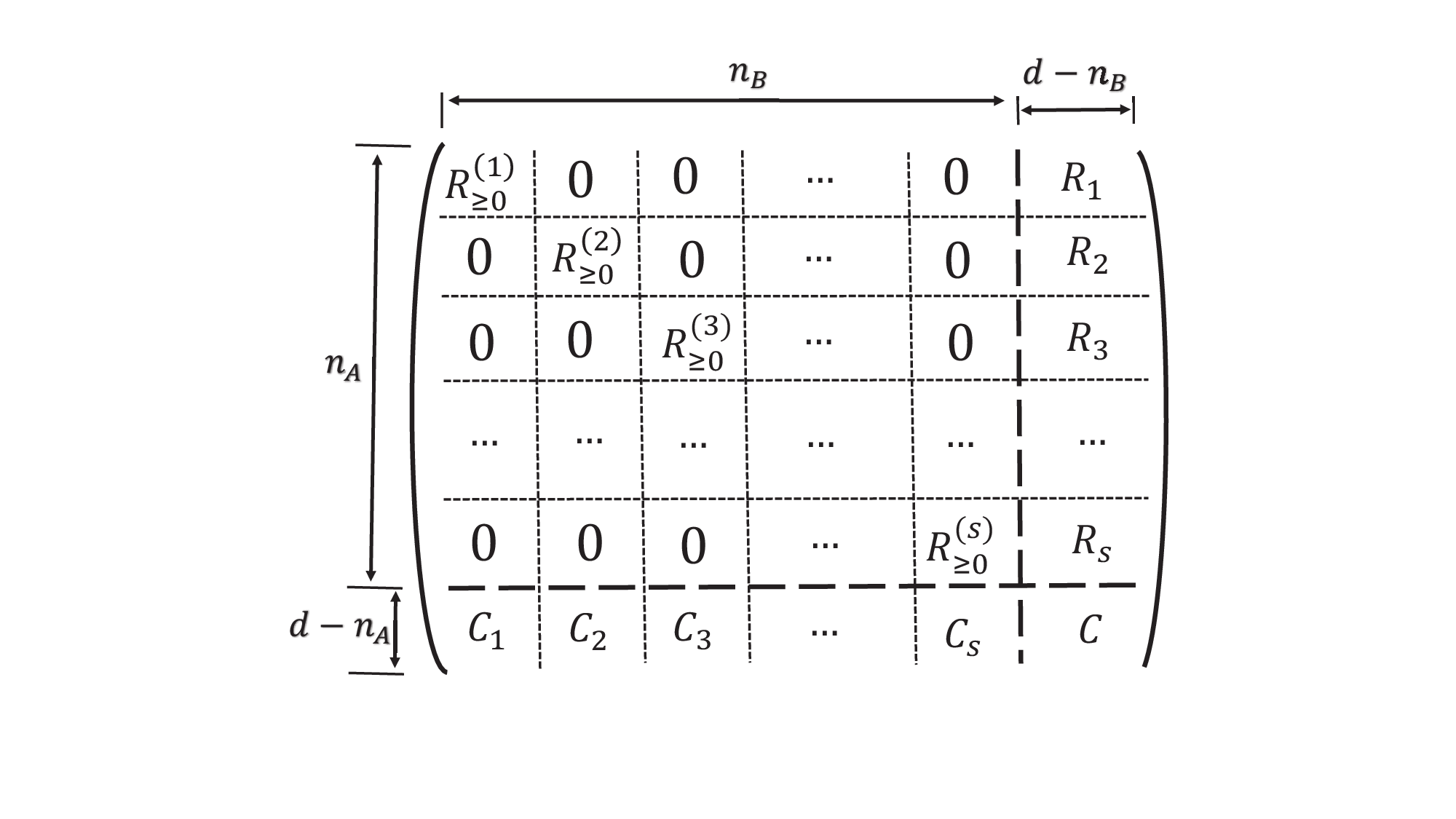}
\caption{Block structure of transition matrix for KD classical pure states.}
\end{figure}
\begin{figure}[!ht]
\includegraphics[width=0.45\textwidth,bb=180 80 800 520,clip]{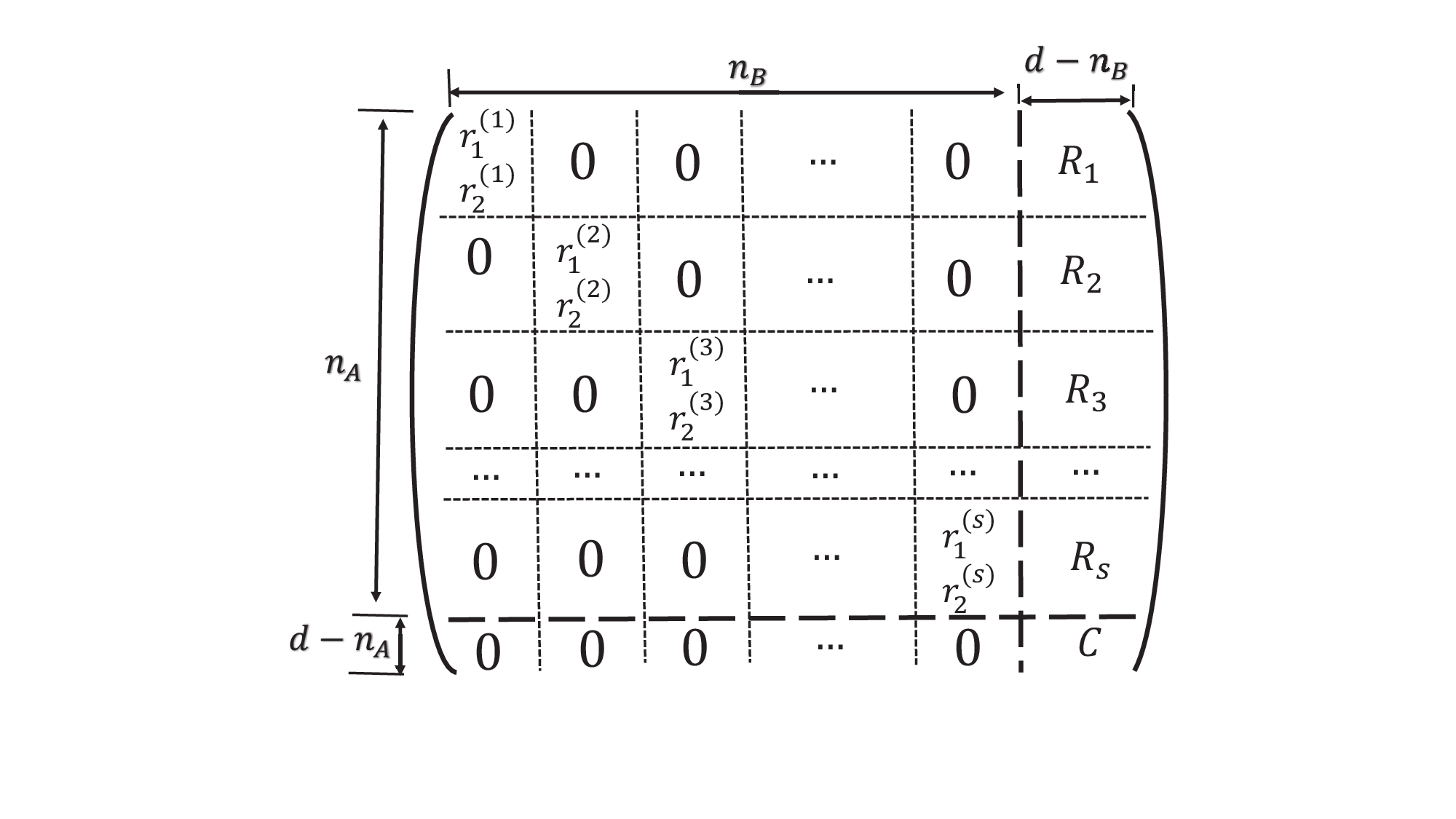}
\caption{Minimum of total number of zeros for given $s$.}
\end{figure}
\begin{figure}[!ht]
\includegraphics[width=0.45\textwidth,bb=180 80 800 520,clip]{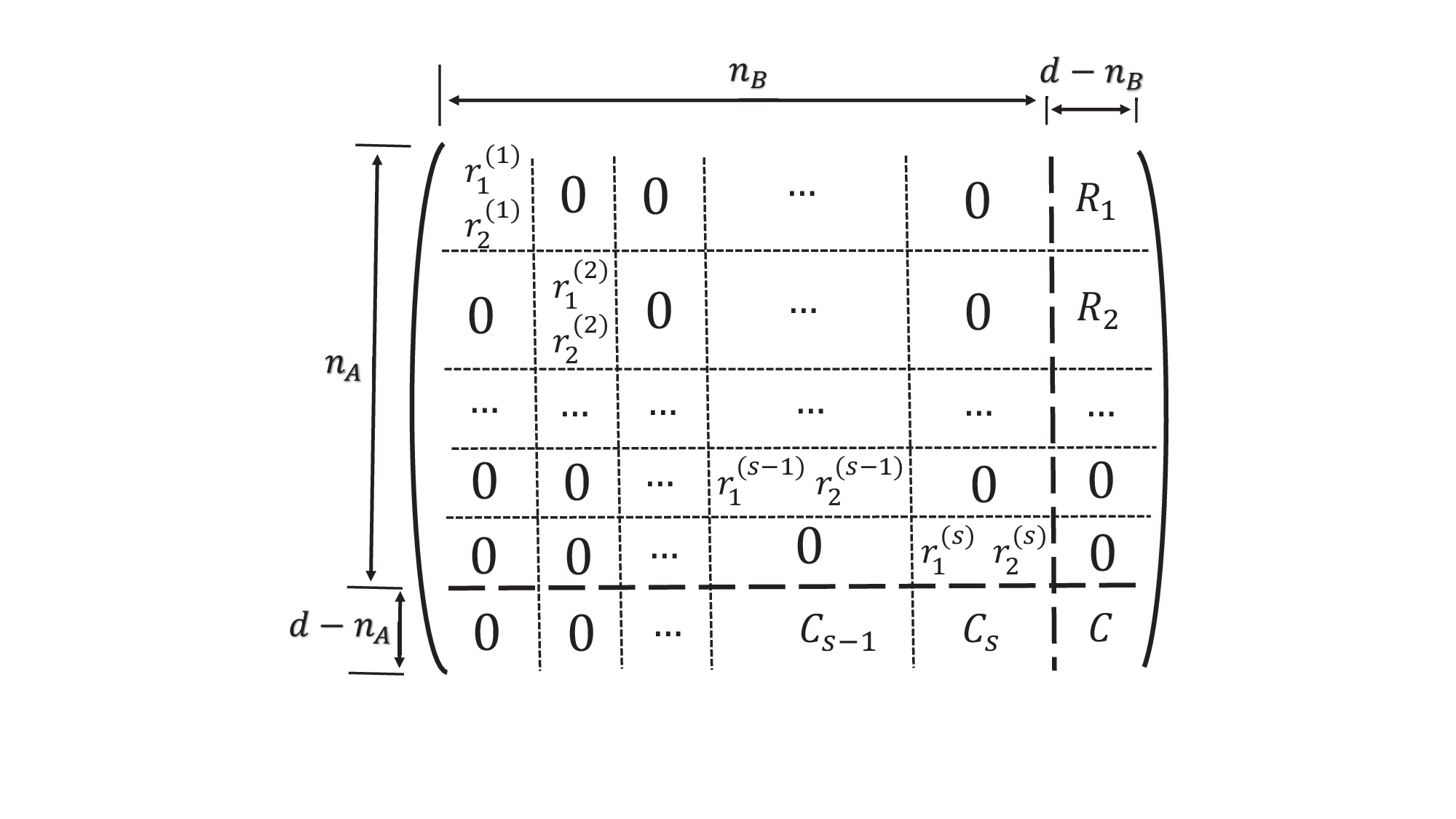}
\caption{Each of $\{R_{\geq 0}^{(j)}\}_{j=1}^{s}$ is either row vector or column vector.}  \label{Fig3}
\end{figure}

\emph{Proof.} \ From Eqs. (\ref{eq2.10},\ref{eq2.11},\ref{eq2.12},\ref{eq2.13}), we have $V=%
\binom{1,2,...,n_{A};}{1,2,...,n_{B}.},$ and
\begin{equation}
\langle \overrightarrow{B}|\overrightarrow{B}\rangle =\langle
\overrightarrow{B}|V^{t}V\overrightarrow{B}\rangle =\langle V\overrightarrow{%
B}|V\overrightarrow{B}\rangle .  \label{eq2.33}
\end{equation}
Since $U^{AB}$ is unitary, then
\begin{equation}
\langle \overrightarrow{B}|\overrightarrow{B}\rangle =\langle U^{AB}%
\overrightarrow{B^{\prime }}|U^{AB}\overrightarrow{B^{\prime }}\rangle
\label{eq2.34}
\end{equation}
with $\overrightarrow{B^{\prime }}=(B_{1},B_{2},...,B_{n_{B}},0,...,0)^{t}$ by appending
zeros for the $\{j\}_{j=n_{B}+1}^{d}$ components. Eqs. (\ref{eq2.33},\ref{eq2.34})
imply that
\begin{equation*}
\binom{n_{A}+1,n_{A}+2,...,d;}{1,2,...,n_{B}.}\overrightarrow{B}=0,
\end{equation*}
i.e.,
\begin{equation}
(C_{1},C_{2},...,C_{s})\overrightarrow{B}=0.  \label{eq2.35}
\end{equation}

Since $U^{AB}$ is unitary, then any column vector in $C_{j}$ is orthogonal to any column vector
in $C_{k}$ when $j\neq k.$ Notice that the components of $\overrightarrow{B}$ are
all positive, then Eq. (\ref{eq2.35}) implies $C_{j}=0$ when cols$(C_{j})=$cols$(R_{\geq 0}^{(j)})=1.$ Also, $R_{\geq 0}^{(j)}\neq 0$ for any $j$, otherwise Eq. (\ref{eq2.35}) implies that the columns containing $C_{j}$ in $U^{AB}$ are linear dependent, which contradicts the unitarity of $U^{AB}$.
$R_{\geq 0}^{(j)}$ must have more than one nonzero element when cols$%
(R_{\geq 0}^{(j)})=1,$ otherwise $U^{AB}$ can be decomposed into the form of direct sum. When cols$(C_{j})=
$cols$(R_{\geq 0}^{(j)})>1,$ Eq. (\ref{eq2.35}) implies that the column
vectors of $C_{j}$ are linearly dependent. Any two distinct column vectors of $U^{AB}$
are orthogonal, and the inner product of any two distinct column vectors of $R_{\geq
0}^{(j)}$ is in $[0,1),$ then the inner product of any two distinct column vectors of
$C_{j}$ is in $(-1,0].$ When cols$(R_{\geq0}^{(j)})\geq 2,$ if $C_{j}$ has a zero column vector, or the column
vectors of $C_{j}$ can be divided into two nonempty sets with orthogonal
spanned spaces, then $R_{\geq 0}^{(j)}$ can be decomposed into the form of direct sum,
this contradicts the assumption that $s$ is the maximum number of blocks. Employing Lemma 1, we see that when cols$(R_{\geq 0}^{(j)})\geq 2,$ the
column vectors of $C_{j}$ (note that all column vectors of $C_{j}$ are not normalized),
 after normalization, must have the form of (A.b) or (A.c). From
Eq. (\ref{eq2.35}) we get that Eq. (\ref{eq2.29}) holds. Eq. (\ref{eq2.30})
similarly holds.

Eq. (\ref{eq2.29}) yields
\begin{eqnarray}
\sum_{j=1}^{s}\text{rank}(C_{j}) &=&\sum_{j=1}^{s}\text{cols}(R_{\geq
0}^{(j)})-s,  \label{eq2.36} \\
\text{rank}(C_{1},C_{2},...,C_{s}) &=&n_{B}-s,  \label{eq2.37}
\end{eqnarray}
where we compute rank$(C_{j})$ by the column rank (recall that for a matrix,
the column rank equals the row rank), and have used the fact that any two
column vectors in distinct $\{C_{j}\}_{j=1}^{s}$ are orthogonal. We compute
rank$(C_{1},C_{2},...,C_{s})$ on the other hand by the row rank, then we get
\begin{equation}
\text{rank}(C_{1},C_{2},...,C_{s})\leq d-n_{A}.  \label{eq2.38}
\end{equation}
Eqs. (\ref{eq2.37},\ref{eq2.38}) certainly result in Eq. (\ref{eq2.31}).

Lastly, we prove Eq. (\ref{eq2.32}). In Fig. 1, we see that
\begin{equation*}
n_{A}(\psi )+n_{B}(\psi )=\sum_{j=1}^{s}(\text{rows}(R_{\geq 0}^{(j)})+\text{%
cols}(R_{\geq 0}^{(j)})).
\end{equation*}
Since each rows$(R_{\geq 0}^{(j)})+$cols$(R_{\geq 0}^{(j)})\geq 3,$ then
\begin{equation*}
n_{A}(\psi )+n_{B}(\psi )\geq 3s.
\end{equation*}
Together with Eq. (\ref{eq1.5}), we certainly get Eq. (\ref{eq2.32}). We
then finished this proof. $\hfill\blacksquare$

We remark that, Eqs. (\ref{eq2.31},\ref{eq2.32}) certainly result in Eq. (\ref{eq1.5}) which was obtained in Ref. \cite{JPA-2021-NYH}. We also remark that, when $s\geq 2,$ from Fig. 1, there must exist two distinct columns
(rows) in $U^{AB}$, such that the number of total zeros in these two distinct columns
(rows) is no less than $n_{A}$ ($n_{B}$). Conversely, if $\max \{n_{A}(\psi ),n_{B}(\psi )\}>\max
\{Z_{r},Z_{c}\},$ with $Z_{c}$ ($Z_{r}$) denoting the maximum number of zeros in any two distinct two columns (rows) in $U^{AB}$, then $s=1$ and Eq. (\ref{eq2.31}) yields $n_{A}(\psi
)+n_{B}(\psi )\leq d+1.$ This returns to the result of Proposition 11 in \cite%
{Bievre-2022-JMP}.

\section{Zeros in transition matrix}

We explore the bounds of $n_{A}(\psi )+n_{B}(\psi )$ for KD classical pure
states in terms of the number of zeros in $U^{AB}.$ To do so, we define $N_{AB}^{(0)}$
as the number of zeros in $U^{AB},$
\begin{equation}
N_{AB}^{(0)}=|\{\langle a_{j}|b_{k}\rangle \Big|\langle a_{j}|b_{k}\rangle
=0,1\leq j,k\leq d\}|.  \label{eq3.1}
\end{equation}
The number of zeros in a unitary matrix is an interesting topic, some recent results are reported in Ref. \cite{Chen-2022-LMA}.
Obviously, $N_{AB}^{(0)}$ keeps invariant if we replace $A=\{|a_{j}\rangle
\}_{j=1}^{d}$ by $\widetilde{A}=\{e^{i\xi _{j}}|a_{j}\rangle \}_{j=1}^{d}$
and replace $B=\{|b_{k}\rangle \}_{k=1}^{d}$ by $\widetilde{B}=\{e^{i\eta
_{k}}|b_{k}\rangle \}_{k=1}^{d},$ where $\{\xi _{j}\}_{j=1}^{d}\subseteq \mathbb{R},$
$\{\eta _{k}\}_{j=1}^{d}\subseteq \mathbb{R}.$ Theorem 3 below provides a link
between $N_{AB}^{(0)}$ and $n_{A}(\psi )+n_{B}(\psi )$ for KD nonclassical
pure states.

\textbf{Theorem 3.} Suppose $|\psi \rangle $ is a KD classical
state, and $U^{AB}$ is as in Fig. 1.
Then it holds that
\begin{equation}
N_{AB}^{(0)}\geq s(2s-1)\text{ for }s\geq 2.  \label{eq3.2}
\end{equation}

\emph{Proof.} For given $s$ with $s\geq 2,$
when $\{R_{\geq 0}^{(j)}\}_{j=1}^{s}$ are all of the column form
$\binom{r_{1}^{(j)}}{r_{2}^{(j)}}$ (or when $\{R_{\geq 0}^{(j)}\}_{j=1}^{s}$ are
all of the row form $(r_{1}^{(j)},r_{2}^{(j)})$ similarly) with $%
r_{1}^{(j)}>0,$ $r_{2}^{(j)}>0,$ $|r_{1}^{(j)}|^{2}+|r_{2}^{(j)}|^{2}=1,$ as
depicted in Fig. 2 and let $d-n_{A}=1,$ then $N_{AB}^{(0)}$ reaches the
minimum. Otherwise, if there exists a $R_{\geq 0}^{(j)}$ with size larger than $\binom{r_{1}^{(j)}}{r_{2}^{(j)}}$ or $(r_{1}^{(j)},r_{2}^{(j)}),$ then deleting
the rows or columns such that $R_{\geq 0}^{(j)}$ shrinks into the form of $\binom{r_{1}^{(j)}}{r_{2}^{(j)}}$ or $(r_{1}^{(j)},r_{2}^{(j)}),$ $N_{AB}^{(0)}$
evidently does not increase in such process. If some of $\{R_{\geq
0}^{(j)}\}_{j=1}^{s}$ are all of the column form $\binom{r_{1}^{(j)}}{r_{2}^{(j)}},$ but others of $\{R_{\geq 0}^{(j)}\}_{j=1}^{s}
$ are all of the row form $(r_{1}^{(j)},r_{2}^{(j)}),$ we can reorder $%
\{R_{\geq 0}^{(j)}\}_{j=1}^{s}$ such that $\{R_{\geq 0}^{(j)}\}_{j=1}^{s_{1}}
$ are all of the column form $\binom{r_{1}^{(j)}}{r_{2}^{(j)}},$ but others $%
\{R_{\geq 0}^{(j)}\}_{j=s_{1}+1}^{s}$ are all of the row form $%
(r_{1}^{(j)},r_{2}^{(j)}),$ as depicted in Fig. 3. In Fig. 3, the number
of zeros in submatrix formed by the columns of $\{R_{\geq
0}^{(j)}\}_{j=1}^{s_{1}}$ and rows of $\{R_{\geq 0}^{(j)}\}_{j=s_{1}+1}^{s}$
is $s_{1}(s-s_{1});$ the number of zeros in submatrix formed by the rows
of $\{R_{\geq 0}^{(j)}\}_{j=1}^{s_{1}}$ and columns of $\{R_{\geq
0}^{(j)}\}_{j=s_{1}+1}^{s}$ is $4s_{1}(s-s_{1}).$ However, in Fig. 2, the
number of zeros in submatrix formed by the columns of $\{R_{\geq
0}^{(j)}\}_{j=1}^{s_{1}}$ and rows of $\{R_{\geq 0}^{(j)}\}_{j=s_{1}+1}^{s}$
is $2s_{1}(s-s_{1});$ the number of zeros in submatrix formed by the rows
of $\{R_{\geq 0}^{(j)}\}_{j=1}^{s_{1}}$ and columns of $\{R_{\geq
0}^{(j)}\}_{j=s_{1}+1}^{s}$ is also $2s_{1}(s-s_{1}).$ Thus $N_{AB}^{(0)}$
in Fig. 3 is greater than $N_{AB}^{(0)}$ in Fig. 2, and Fig. 2 reaches
the minimum of $N_{AB}^{(0)}.$

Apparently, in Fig. 2, $N_{AB}^{(0)}=s(2s-1)$ when $s\geq 2$, this certainly yields Eq. (\ref{eq3.2}). We then finished this proof.
$\hfill\blacksquare$

With Theorem 2 and Theorem 3, we now establish Theorem 4 which provides a sufficient condition for pure KD nonclassial states.

\textbf{Theorem 4.} For the transition matrix $U^{AB}$ and pure state $|\psi \rangle $, if
\begin{eqnarray}
2\leq \frac{1+\sqrt{1+8N_{AB}^{(0)}}}{4}<n_{A}(\psi )+n_{B}(\psi )-d  \label{eq3.3}
\end{eqnarray}
or
\begin{eqnarray}
\frac{1+\sqrt{1+8N_{AB}^{(0)}}}{4}<2\leq n_{A}(\psi )+n_{B}(\psi )-d,  \label{eq3.4}
\end{eqnarray}
then $|\psi \rangle $ is KD nonclassical.

\emph{Proof.}
Theorem 4 is a result of Eqs. (\ref{eq2.31},\ref{eq3.2}). We prove the contrapositive of Theorem 4. Suppose pure state $|\psi \rangle $ is KD classical, then we write the transition matrix $U^{AB}$ as in Fig. 1. If $s=1$, then Eq. (\ref{eq2.31}) yields $n_{A}(\psi )+n_{B}(\psi )-d\leq 1$, and both Eqs. (\ref{eq3.3},\ref{eq3.4}) do not hold.

If $s>1$, then both Eqs. (\ref{eq2.31},\ref{eq3.2}) hold.  Eq. (\ref{eq3.2}) implies
\begin{eqnarray}
\frac{1+\sqrt{1+8N_{AB}^{(0)}}}{4}\geq s,  \label{eqR.1}
\end{eqnarray}
and Eq. (\ref{eq2.31}) implies
\begin{eqnarray}
n_{A}(\psi )+n_{B}(\psi )-d \leq s.    \label{eqR.2}
\end{eqnarray}
Eqs. (\ref{eqR.1},\ref{eqR.2}) implies
\begin{eqnarray}
n_{A}(\psi )+n_{B}(\psi )-d \leq s \leq \frac{1+\sqrt{1+8N_{AB}^{(0)}}}{4}.    \label{eqR.3}
\end{eqnarray}
Eq. (\ref{eqR.3}) obviously contradicts both Eqs. (\ref{eq3.3},\ref{eq3.4}).
$\hfill\blacksquare$

For a special case in Theorem 4, we have Corollary 2 below.

\textbf{Corollary 2.} For the transition matrix $U^{AB}$ and pure state $|\psi \rangle $, if
\begin{eqnarray}
N_{AB}^{(0)} &<&6,  \label{eq3.5} \\
n_{A}(\psi )+n_{B}(\psi ) &>&d+1,  \label{eq3.6}
\end{eqnarray}
then $|\psi \rangle $ is KD nonclassical.

\emph{Proof.}
We can directly check that Eqs. (\ref{eq3.5},\ref{eq3.6}) lead to Eq. (\ref{eq3.4}).
$\hfill\blacksquare$

We remark that Corollary 2 improved the Theorem 4 in Ref. \cite%
{PRL-2021-Bievre} where Eq. (\ref{eq3.5}) is replaced by $N_{AB}^{(0)}=0.$

\section{Examples}

In this section, we provide some examples to demonstrate the applications of
the results in section II and section III.

\emph{Example 1.} The transition matrix is $U_{5}$ expressed in
Fig. 4.

\begin{widetext}
\begin{figure}[!ht]
\includegraphics[width=0.8\textwidth,bb=0 80 1000 520,clip]{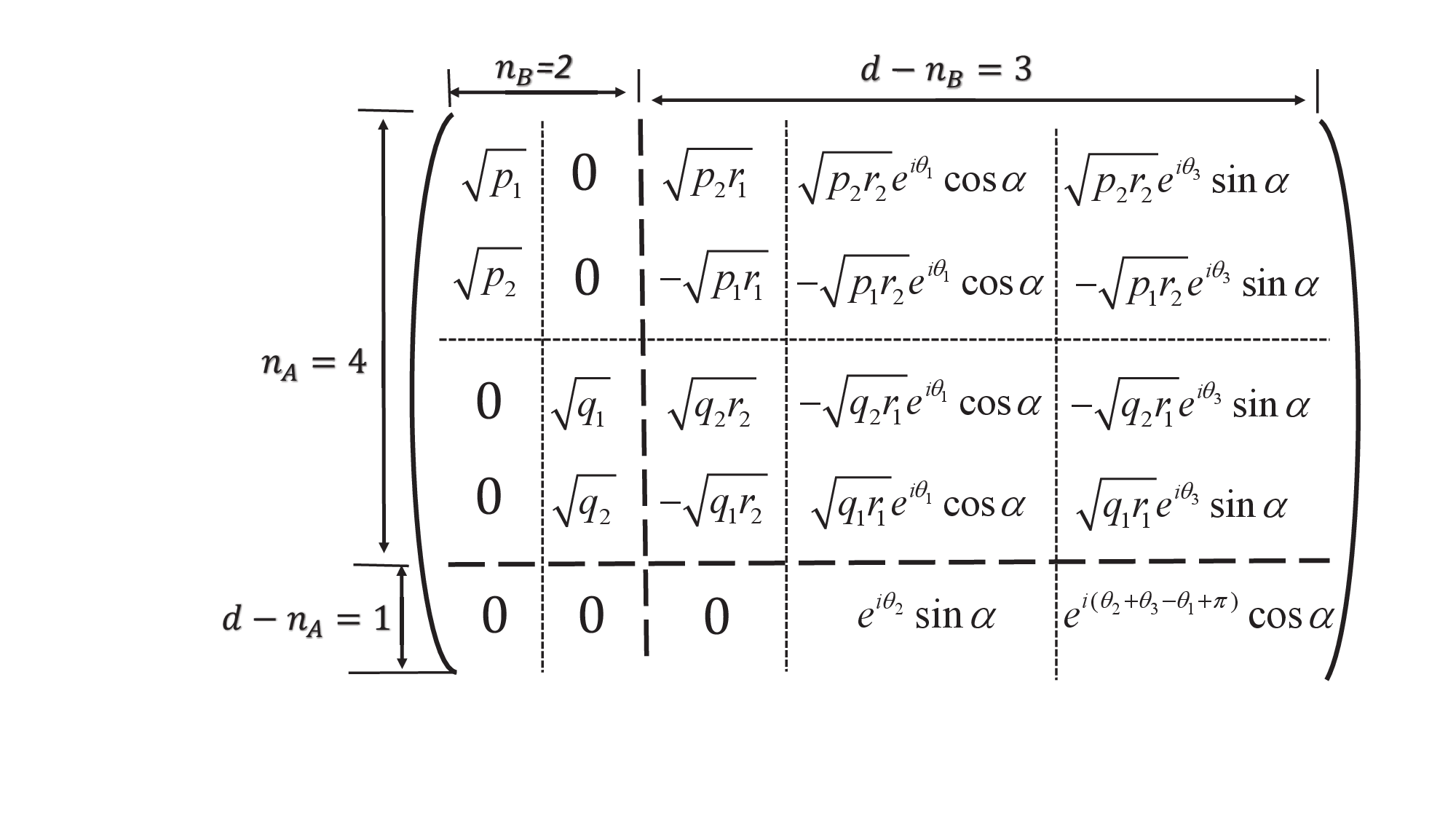}
\captionsetup{width=0.9\textwidth}
\caption{Block structure of transition matrix $U_{5}$ for KD classical pure state in \ref{eq4.1}.}
\end{figure}
\end{widetext}

In $U_{5},$ $d=5,$ $\{p_{1},q_{1},r_{1}\}\subseteq (0,1),$ $p_{2}=1-p_{1},$ $%
q_{2}=1-q_{1},$ $r_{2}=1-r_{1},$ $\alpha \in (0,\frac{\pi }{2}),$ $\{\theta
_{1},\theta _{2},\theta _{3}\}\subseteq \mathbb{R}.$ We see that
$U_{5}$ can not be written in the form of direct sum. Consider the pure
state expressed in the basis $B=\{|b_{k}\rangle \}_{k=1}^{5}$ as
\begin{equation}
|\psi \rangle =\sqrt{t_{1}}|b_{1}\rangle +\sqrt{t_{2}}|b_{2}\rangle ,  \label{eq4.1}
\end{equation}
with $t_{1}\in (0,1),$ $t_{1}+t_{2}=1.$

Direct computation shows that $|\psi \rangle $ is expressed in the basis $%
A=\{|a_{j}\rangle \}_{j=1}^{5}$ as
\begin{equation*}
|\psi \rangle =\sqrt{t_{1}p_{1}}|a_{1}\rangle +\sqrt{t_{1}p_{2}}%
|a_{2}\rangle +\sqrt{t_{2}q_{1}}|a_{3}\rangle +\sqrt{t_{2}q_{2}}%
|a_{4}\rangle .
\end{equation*}
Hence $n_{B}(\psi )=2,$ $n_{A}(\psi )=4,$ and $s=2$ as shown in Fig. 4. From
Eq. (\ref{eq1.1}) we can directly get that
\begin{equation*}
Q(|\psi \rangle)=\left(
\begin{array}{ccccc}
t_{1}p_{1} & 0 & 0 & 0 & 0 \\
t_{1}p_{2} & 0 & 0 & 0 & 0 \\
0 & t_{2}q_{1} & 0 & 0 & 0 \\
0 & t_{2}q_{2} & 0 & 0 & 0 \\
0 & 0 & 0 & 0 & 0%
\end{array}%
\right) .
\end{equation*}
That is, $|\psi \rangle $ is KD classical. We can check that Eqs. (\ref{eq2.29},\ref{eq2.30},\ref{eq2.31},\ref{eq2.32}) all
holds for $U_{5}$ and $|\psi \rangle .$

Further, in $U_{5},$ $N_{AB}^{(0)}=7.$ We can check that Theorem 3
holds for $U_{5}$ and $|\psi \rangle .$ Theorem 4 implies that for the
pure state $|\varphi \rangle $ if $n_{B}(\varphi )+$ $n_{A}(\varphi )>d+2=7$
then $|\varphi \rangle $ is KD nonclassical.

\emph{Example 2.} The transition matrix is discrete Fourier transformation
(DFT).

De Bi\`{e}vre \cite{PRL-2021-Bievre} conjectured that for DFT of dimension $%
d,$ a pure state $|\psi \rangle $ is KD classical iff $n_{A}(\psi
)n_{B}(\psi )=d.$ We see that Corollary 1 answered this conjecture in the
affirmative. When $d=5$ and $d=6,$ these results return to (a) and (b) of
Figure 1 in Ref. \cite{PRL-2021-Bievre}; when $d=7$ these results return to the right panel of
Figure 1 in Ref. \cite{Bievre-2022-JMP}.

\emph{Example 3. }The transition matrix is Tao matrix.

The Tao matrix is the following unitary matrix $U_{T}$ with $d=6,$
\begin{eqnarray}
U_{T}=\frac{1}{\sqrt{6}}\left(
\begin{array}{cccccc}
1 & 1 & 1 & 1 & 1 & 1 \\
1 & 1 & \omega & \omega & \omega ^{2} & \omega ^{2} \\
1 & \omega & 1 & \omega ^{2} & \omega ^{2} & \omega \\
1 & \omega & \omega ^{2} & 1 & \omega & \omega ^{2} \\
1 & \omega ^{2} & \omega ^{2} & \omega & 1 & \omega \\
1 & \omega ^{2} & \omega & \omega ^{2} & \omega & 1%
\end{array}%
\right) ,   \nonumber
\end{eqnarray}
with $\omega =\exp \left( i\frac{2\pi }{3}\right).$

Apparently, for Tao matrix, bases $A$ and $B$ are MUBs. From Fact 3 we see
that there exist pure KD classical states (basis states in $A$) satisfying $%
(n_{A}=1$, $n_{B}=6)$ and there exist pure KD classical states (basis states
in $B$) satisfying $(n_{A}=6$, $n_{B}=1).$ With Corollary 1, the other
possible pure KD classical states exist only when $\{n_{A},n_{B}\}=\{2,3\}.$
It is proved that \cite{Bievre-2022-JMP} there does not exist pure state satisfying $%
\{n_{A},n_{B}\}=\{2,3\}$. As a result, for Tao matrix, the only pure KD classical states are
basis states.

\emph{Example 4. }The transition matrix is
\begin{equation*}
U_{4}=\frac{1}{\sqrt{2}}\left(
\begin{array}{cccc}
1 & 1 & 0 & 0 \\
0 & 0 & 1 & 1 \\
\cos \alpha  & -\cos \alpha  & \sin \alpha  & -\sin \alpha  \\
\sin \alpha  & -\sin \alpha  & -\cos \alpha  & \cos \alpha
\end{array}%
\right) ,
\end{equation*}
with $\alpha \in (0,\frac{\pi }{2}).$

Evidently, $U_{4}$ can not be written in the form of direct sum. In $U_{4}$,
$d=4$, $N_{AB}^{(0)}=4.$ Theorem 4 yields that the pure state $%
|\psi \rangle $ is KD nonclassical if $n_{A}(\psi )+n_{B}(\psi )>d+1=5.$ We
give an explicit example. Suppose the pure state $|\psi \rangle $ is
expressed in the basis $B=\{|b_{k}\rangle \}_{k=1}^{4}$ as
\begin{equation}
|\psi \rangle =\sqrt{t_{1}}|b_{1}\rangle +\sqrt{t_{3}}|b_{3}\rangle ,  \label{eq4.2}
\end{equation}
with $t_{1}\in (0,1),$ $t_{1}+t_{3}=1,$ $\sqrt{\frac{t_{3}}{t_{1}}}\neq \tan
\alpha .$
Then direct computation shows that $|\psi \rangle $ is expressed in the
basis $A=\{|a_{j}\rangle \}_{j=1}^{5}$ as
\begin{eqnarray*}
|\psi \rangle  &=&\frac{1}{\sqrt{2}}[\sqrt{t_{1}}|a_{1}\rangle +\sqrt{t_{3}}%
|a_{2}\rangle  \\
&&+(\sqrt{t_{1}}\cos \alpha +\sqrt{t_{3}}\sin \alpha )|a_{3}\rangle  \\
&&+(\sqrt{t_{1}}\sin \alpha -\sqrt{t_{3}}\cos \alpha )|a_{4}\rangle ].
\end{eqnarray*}
Hence $n_{B}(\psi )=2,$ $n_{A}(\psi )=4.$ From Eq. (\ref{eq1.1}) we directly get that
\begin{eqnarray*}
Q_{41}(\psi ) &=&\frac{1}{2}\sqrt{t_{1}}(\sqrt{t_{1}}\sin \alpha -\sqrt{t_{3}%
}\cos \alpha )\sin \alpha , \\
Q_{43}(\psi ) &=&\frac{-1}{2}\sqrt{t_{3}}(\sqrt{t_{1}}\sin \alpha -\sqrt{%
t_{3}}\cos \alpha )\cos \alpha .
\end{eqnarray*}
Consequently, one of $\{Q_{51}(\psi )$,$Q_{52}(\psi )\}$ must be negative,
and $|\psi \rangle $ is indeed KD nonclassical.

Remark that, in $U_{4}$, $Z_{r}=4,$ $Z_{c}=2$, the conditions of Proposition
11 or Theorem 12 in Ref. \cite{Bievre-2022-JMP} are not satisfied, thus we can not apply
Proposition 11 or Theorem 12 in Ref. \cite{Bievre-2022-JMP} to $U_{4}.$ Example 4 then
shows that for some states Theorem 4 is more advantageous than Proposition
11 or Theorem 12 in Ref. \cite{Bievre-2022-JMP}.

\emph{Example 5. }The transition matrix is $U_{6}$ or $U_{6}^{\prime }$ as
\begin{eqnarray}
U_{6}=\frac{1}{\sqrt{5}}\left(
\begin{array}{cccccc}
0 & 1 & 1 & 1 & 1 & 1 \\
1 & 0 & 1 & -1 & 1 & -1 \\
1 & 1 & 0 & -1 & -1 & 1 \\
1 & -1 & -1 & 0 & 1 & 1 \\
1 & 1 & -1 & 1 & 0 & -1 \\
1 & -1 & 1 & 1 & -1 & 0%
\end{array}%
\right) , \nonumber \\
U_{6}^{\prime }=\frac{1}{\sqrt{5}}\left(
\begin{array}{cccccc}
0 & 1 & 1 & 1 & 1 & 1 \\
1 & 0 & 1 & -1 & 1 & -1 \\
1 & 1 & 0 & 1 & -1 & -1 \\
1 & -1 & 1 & 0 & -1 & 1 \\
1 & 1 & -1 & -1 & 0 & 1 \\
1 & -1 & -1 & 1 & 1 & 0%
\end{array}%
\right) .   \nonumber
\end{eqnarray}
In $U_{6}$ or $U_{6}^{\prime }$, $d=6$, $N_{AB}^{(0)}=6.$ Eq. (\ref{eq3.3}) in Theorem
4 yields that the pure state $|\psi \rangle $ is KD nonclassical if $%
n_{A}(\psi )+n_{B}(\psi )>8.$ In $U_{6}$ or $U_{6}^{\prime }$, $%
Z_{r}=Z_{c}=2.$ Proposition 11 in Ref. \cite{Bievre-2022-JMP} implies that if $\max
\{n_{A}(\psi ),n_{B}(\psi )\}>2$ and $n_{A}(\psi )+n_{B}(\psi )>7,$ then is
KD nonclassical. Theorem 12 in Ref. \cite{Bievre-2022-JMP} implies that if $n_{A}(\psi
)+n_{B}(\psi )>7,$ then $|\psi \rangle $ is KD nonclassical.  Example 5 then shows that for
some states Proposition 11 or Theorem 12
in Ref. \cite{Bievre-2022-JMP} is more advantageous than Theorem 4.

\section{Summary}

We established general structure for KD classical pure states in Theorem 1
and Theorem 2. We explored the links between KD classical pure states and the
number of zeros in transition matrix in Theorem 3 and Theorem 4. Also, we provide some examples to demonstrate the applications of our results.

We emphasize that whether a pure state is KD classical is dependent on the choice of orthonormal bases $A$ and $B.$ For example, if we choose $A=B,$ then Eq. (\ref{eq1.1}) shows that any pure states are KD classical. Conversely, for any orthonormal bases $A$ and $B,$ Fact 3 implies that there must exist some KD classical pure states. The choice of orthonormal bases $A$ and $B$ would be specific depending on the concrete task in application, for example, the MUBs or DFT discussed in section IV.

There remained many open questions after this paper. First, how to apply the
results of this paper to relevant experimental scenarios. Second, how to
generalize the results of this paper to mixed states. Third, how to find out
the KD classical (pure and mixed) states for some special transition
matrices.

\section*{ACKNOWLEDGMENTS}
This work was supported by the Natural Science Basic Research Plan in
Shaanxi Province of China (Program No. 2022JM-012). The author thanks Stephan De Bi\`{e}vre and Kailiang Lin for insightful discussions. The author thanks Tao Li for reading the manuscript. The author also thanks the anonymous referees for constructive comments. After completing this work, I became aware of the recent work \cite{Bievre-2023-arXiv} which discussed the structure of Kirkwood-Dirac classical mixed states.

\section*{Appendix: a mathematical result}
\setcounter{equation}{0} \renewcommand%
\theequation{A\arabic{equation}}

We prove a mathematical result in Lemma 1 below, which is useful in the proof of Theorem 2.

\textbf{Lemma 1.} Let $\{\overrightarrow{v_{j}}\}_{j=1}^{n}\subseteq
\mathbb{C}^{d}\backslash \{0\}$ and $||\overrightarrow{v_{j}}||=1$ for all $j.$ If
\begin{equation}
-1\leq \langle \overrightarrow{v_{j}}|\overrightarrow{v_{k}}\rangle \leq 0
\label{eq2.18}
\end{equation}
for all $1\leq j<k\leq n$, then $\{\overrightarrow{v_{j}}\}_{j=1}^{n}$ has
the unique decomposition
\begin{equation}
\{\overrightarrow{v_{j}}\}_{j=1}^{n}=\cup _{\alpha =1}^{\widetilde{n}%
}S_{\alpha },  \label{eq2.19}
\end{equation}
where $S_{\alpha }\neq \varnothing $ for any $\alpha ,$ $S_{\alpha }\cap
S_{\beta }=\varnothing $ for any $\alpha \neq \beta ,$ $\langle
\overrightarrow{v_{j}}|\overrightarrow{v_{k}}\rangle =0$ for any $%
\overrightarrow{v_{j}}\in S_{\alpha }$, $\overrightarrow{v_{k}}\in S_{\beta
} $ and $\alpha \neq \beta ,$ any $S_{\alpha }$ has one of the structures
(A.a), (A.b), or (A.c) below, and any $S_{\alpha }$ can not be further
decomposed as $S_{\alpha }=S_{\alpha _{1}}\cup S_{\alpha _{2}}$ with $%
\{S_{\alpha _{1}},S_{\alpha _{2}}\}$ having similar properties as $S_{\alpha
}.$

(A.a) $S_{a}=\{\overrightarrow{v_{j}}\}.$ $S_{a}$ contains only one element.

(A.b) $S_{b}=\{\overrightarrow{v_{j}},\overrightarrow{v_{k}}\},$ $%
\overrightarrow{v_{j}}=-\overrightarrow{v_{k}}.$ $S_{b}$ contains two
elements with opposite directions.

(A.c) $S_{c}=\{\overrightarrow{v_{j}}\}_{j},$
\begin{equation}
-1<\langle \overrightarrow{v_{j}}|\overrightarrow{v_{k}}\rangle \leq 0\text{
for } \forall \overrightarrow{%
v_{j}},\overrightarrow{v_{k}} \in S_{c}, \overrightarrow{v_{j}}\neq \overrightarrow{v_{k}}.  \label{eq2.20}
\end{equation}
$S_{c}$ contains more than one element.

For the case of (A.a), $\dim ($span$\{S_{a}\})=1;$ for the case of (A.b), $%
\dim ($span$\{S_{b}\})=1;$ for the case of (A.c),
\begin{equation}
|S_{c}|=\dim (\text{span}\{S_{c}\})\text{ or }|S_{c}|=\dim (\text{span}%
\{S_{c}\})+1.  \label{eq2.21}
\end{equation}

\emph{Example 6.} We first give a concrete example to explain Lemma 1. Suppose $\{\overrightarrow{v_{j}}\}_{j=1}^{6}\subseteq
\mathbb{C}^{4}\backslash \{0\}$ and $||\overrightarrow{v_{j}}||=1$ for all $j,$  and in a fixed orthonormal basis of $\mathbb{C}^{4},$ $\{\overrightarrow{v_{j}}\}_{j=1}^{6}$ read
\begin{eqnarray*}
\overrightarrow{v_{1}}&=&(1,0,0,0), \\
\overrightarrow{v_{2}}&=&(0,1,0,0), \\
\overrightarrow{v_{3}}&=&(0,-1,0,0), \\
\overrightarrow{v_{4}}&=&(0,0,-\frac{1}{\sqrt{2}},-\frac{1}{\sqrt{2}}), \\
\overrightarrow{v_{5}}&=&(0,0,1,0), \\
\overrightarrow{v_{6}}&=&(0,0,0,1).
\end{eqnarray*}
We see that $\{\overrightarrow{v_{j}}\}_{j=1}^{6}$ satisfies Eq. (\ref{eq2.18}). Eq. (\ref{eq2.19}) yields $\{\overrightarrow{v_{j}}\}_{j=1}^{6}=S_{1} \cup S_{2} \cup S_{3},$ with $S_{1}=\{\overrightarrow{v_{1}}\},$ $S_{2}=\{\overrightarrow{v_{2}},\overrightarrow{v_{3}}\},$ $S_{3}=\{\overrightarrow{v_{4}},\overrightarrow{v_{5}},\overrightarrow{v_{6}}\}.$ For this case, $S_{1}$ has the form of (A.a), $S_{2}$ has the form of (A.b), $S_{3}$ has the form of (A.c).

Next we consider $\{\overrightarrow{v_{j}}\}_{j=1}^{5}.$  We see that $\{\overrightarrow{v_{j}}\}_{j=1}^{5}$ satisfies Eq. (\ref{eq2.18}). Eq. (\ref{eq2.19}) yields $\{\overrightarrow{v_{j}}\}_{j=1}^{5}=S_{1} \cup S_{2} \cup S'_{3},$ with $S'_{3}=\{\overrightarrow{v_{4}},\overrightarrow{v_{5}}\}.$ For this case, $S'_{3}$ has the form of (A.c).

We depict  $S_{1},$  $S_{2}$, $S_{3}$ and $S'_{3}$ in Fig. 5. Evidently,
\begin{eqnarray*}
|S'_{3}|&=&\dim (\text{span}\{S'_{3}\})=2, \\
|S_{3}|&=&\dim (\text{span}\{S_{3}\})+1=3.
\end{eqnarray*}

\begin{figure}[!ht]
\includegraphics[width=0.45\textwidth,bb=75 50 700 520,clip]{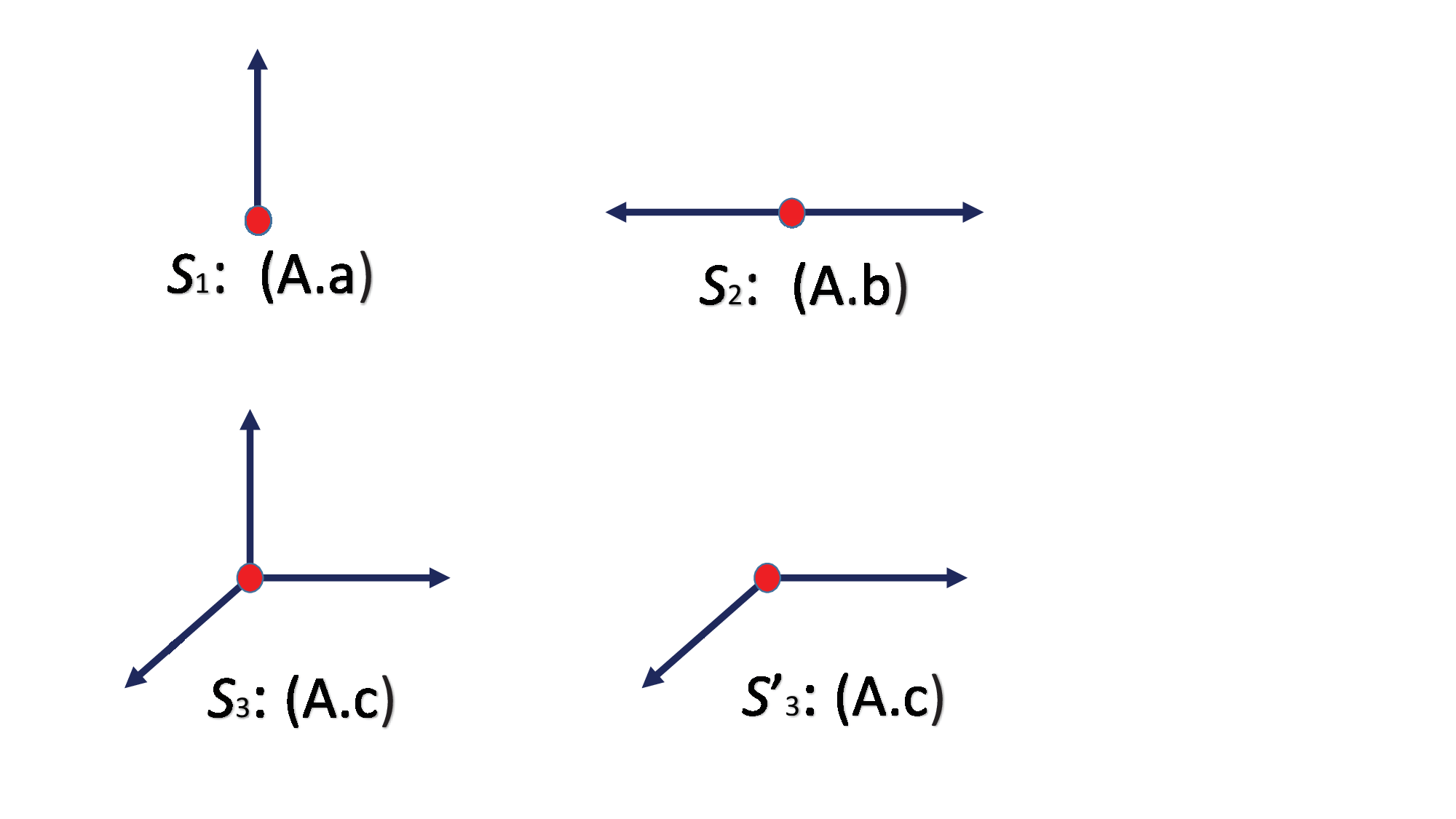}
\caption{In Example 6, $S_{1}$ has the form of (A.a), $S_{2}$  has the form of (A.b), $S_{3}$ and $S'_{3}$ have the form of (A.c).}
\end{figure}

\emph{Proof of Lemma 1.}
For $d=1,$ since $\{\overrightarrow{v_{j}}\}_{j=1}^{n}\subseteq
\mathbb{C}^{d}\backslash \{0\}$ satisfies $||\overrightarrow{v_{j}}||=1$ for all $j$ and Eq. (\ref{eq2.18}), then it must hold that $\{\overrightarrow{v_{j}}\}_{j=1}^{n}=\{\overrightarrow{v_{1}}\}$ or $\{\overrightarrow{v_{j}}\}_{j=1}^{n}=\{\overrightarrow{v_{1}},-\overrightarrow{v_{1}}\}$, then  $\widetilde{n}=1,$ the claim obviously holds.

For $d\geq 2,$ we can always divide $\{\overrightarrow{v_{j}}\}_{j=1}^{n}$
into a union of nonempty sets $\cup _{\alpha =1}^{\widetilde{n}}S_{\alpha },$
such that any two vectors in distinct $\{S_{\alpha }\}_{\alpha =1}^{%
\widetilde{n}}$ are orthogonal, and any $S_{\alpha }$ can not be so
decomposed further. If there exists $\{\overrightarrow{v_{1}}\}\subseteq \{%
\overrightarrow{v_{j}}\}_{j=1}^{n}$ such that $\langle \overrightarrow{v_{j}}%
|\overrightarrow{v_{1}}\rangle =0$ for any $j\geq 2,$ let $\{\overrightarrow{%
v_{1}}\}$ be one element in the decomposition of Eq. (\ref{eq2.19}). If
there exists $\{\overrightarrow{v_{1}},\overrightarrow{v_{2}}\}\subseteq \{%
\overrightarrow{v_{j}}\}_{j=1}^{n}$ such that $\overrightarrow{v_{1}}=-%
\overrightarrow{v_{2}},$ then Eq. (\ref{eq2.18}) implies that $\langle
\overrightarrow{v_{j}}|\overrightarrow{v_{1}}\rangle =-\langle
\overrightarrow{v_{j}}|\overrightarrow{v_{2}}\rangle =0$ for any $j>2,$ let $%
\{\overrightarrow{v_{1}},\overrightarrow{v_{2}}\}$ be one element in the
decomposition of Eq. (\ref{eq2.19}). In this way, we can take out all such
elements of structure (A.a) and structure (A.b) in the decomposition of Eq. (%
\ref{eq2.19}). The remaining elements in Eq. (\ref{eq2.19}) must satisfy Eq.
(\ref{eq2.20}), and we only need to prove Eq. (\ref{eq2.21}).

We prove Eq. (\ref{eq2.21}) by induction in the dimension $\dim ($span$%
\{S_{c}\})$. When $\dim ($span$\{S_{c}\})=2,$ under any orthonormal basis $\{\overrightarrow{e_{j}}\}_{j=1}^{2}$
of span$\{S_{c}\},$ let $\overrightarrow{v_{j}}=\overrightarrow{e_{j}}$ for $%
\{j\}_{j=1}^{2},$ $\overrightarrow{v_{3}}=-\frac{1}{\sqrt{2}}\sum_{j=1}^{2}%
\overrightarrow{e_{j}}.$ These three normalized vectors $\{v_{1},v_{2},v_{3}\}$ certainly satisfy $-1<\langle \overrightarrow{v_{j}}|%
\overrightarrow{v_{k}}\rangle \leq 0$ for all $1\leq j<k\leq 3$. Let $S_{c}=\{v_{1},v_{3}\},$
then $|S_{c}|=\dim (\text{span}\{S_{c}\}).$  Let $S_{c}=\{v_{1},v_{2},v_{3}\},$ then $|S_{c}|=\dim (\text{span}\{S_{c}\})+1.$
We now show that there do not exist four
normalized vectors satisfying $-1<\langle \overrightarrow{v_{j}}|%
\overrightarrow{v_{k}}\rangle \leq 0$ for all $1\leq j<k\leq 4$. Otherwise,
if there exist such four normalized vectors $\{\overrightarrow{v_{j}}%
\}_{j=1}^{4},$ then we can choose an orthonormal basis $\{\overrightarrow{%
e_{j}}\}_{j=1}^{2}$ of span$\{S_{c}\}$ with $\overrightarrow{e_{1}}=$ $%
\overrightarrow{v_{1}},$ and expand $\{\overrightarrow{v_{j}}\}_{j=1}^{4}$
in $\{\overrightarrow{e_{j}}\}_{j=1}^{2}$ as
\begin{eqnarray*}
\overrightarrow{v_{1}} &=&(1,0), \\
\overrightarrow{v_{2}} &=&(x_{21},x_{22}), \\
\overrightarrow{v_{3}} &=&(x_{31},x_{32}), \\
\overrightarrow{v_{4}} &=&(x_{41},x_{42}).
\end{eqnarray*}

Applying Eq. (\ref{eq2.20}) to $\{\langle \overrightarrow{v_{1}}|%
\overrightarrow{v_{j}}\rangle \}_{j=2}^{4},$ we see that $%
\{x_{21},x_{31},x_{41}\}$ are all real and are all in $(-1,0].$ Applying Eq.
(\ref{eq2.20}) to $\{\langle \overrightarrow{v_{j}}|\overrightarrow{v_{k}}%
\rangle \}_{2\leq j<k\leq 4}$, we see that $\{x_{22}x_{32}^{*},x_{22}x_{42}^{*},x_{32}x_{42}^{*}\}$ are all real, thus there exist $\{\theta,x_{22}',x_{32}',x_{42}'\}\subseteq \mathbb{R}$ such that $\{x_{22}=x_{22}'e^{i\theta},x_{32}=x_{32}'e^{i\theta},x_{42}=x_{42}'e^{i\theta}\}$. There is at most one zero in $\{x_{21},x_{31},x_{41}\}$,
otherwise, suppose for example $x_{21}=x_{31}=0,$ then $%
|x_{22}|=|x_{32}|=1=|x_{22}x_{32}|=|\langle \overrightarrow{v_{2}}|%
\overrightarrow{v_{3}}\rangle |$, contradicting Eq. (\ref{eq2.20}).
Similarly, there is no zero in $\{x_{22},x_{32},x_{42}\}$. Then Eq. (\ref%
{eq2.20}) requires $x_{22}'x_{32}'\leq 0,$ $x_{32}'x_{42}'\leq 0$ and $%
x_{42}'x_{22}'\leq 0$ all hold, these are certainly impossible. Hence, there
do not exist four normalized vectors $\{\overrightarrow{v_{j}}\}_{j=1}^{4}$
satisfying Eq. (\ref{eq2.20}), and then Eq. (\ref{eq2.21}) holds for $\dim ($%
span$\{S_{c}\})=2.$

Suppose Eq. (\ref{eq2.21}) is true for all space dimensions $\dim ($span$%
\{S_{c}\})=1,2,3,..,D-1$, with $D\geq 3.$ Now for the space dimension $\dim
( $span$\{S_{c}\})=D,$ there indeed exist $D+1$ normalized vectors
satisfying $-1<\langle \overrightarrow{v_{j}}|\overrightarrow{v_{k}}\rangle
\leq 0$ for all $1\leq j<k\leq D+1.$ For example, under any orthonormal
basis $\{\overrightarrow{e_{j}}\}_{j=1}^{D}$ of span$\{S_{c}\},$ let $%
\overrightarrow{v_{j}}=\overrightarrow{e_{j}}$ for $\{j\}_{j=1}^{D},$ $%
\overrightarrow{v_{D+1}}=-\frac{1}{\sqrt{D}}\sum_{j=1}^{D}\overrightarrow{%
e_{j}}.$   Let $S_{c}=\{v_{j}\}_{j=2}^{D+1},$
then $|S_{c}|=\dim (\text{span}\{S_{c}\}).$  Let  $S_{c}=\{v_{j}\}_{j=1}^{D+1},$ then $|S_{c}|=\dim (\text{span}\{S_{c}\})+1.$

Suppose there exist $D+2$ normalized vectors $\{\overrightarrow{%
v_{j}}\}_{j=1}^{D+2}$ satisfying Eq. (\ref{eq2.20}). Then we can choose an
orthonormal basis $\{\overrightarrow{e_{j}}\}_{j=1}^{D}$ of span$\{S_{c}\}$
with $\overrightarrow{e_{1}}=$ $\overrightarrow{v_{1}},$ and expand $\{%
\overrightarrow{v_{j}}\}_{j=1}^{D+2}$ in $\{\overrightarrow{e_{j}}%
\}_{j=1}^{D}$ as
\begin{eqnarray*}
\overrightarrow{v_{1}} &=&(1,0,0,...,0), \\
\overrightarrow{v_{2}} &=&(x_{21},x_{22},x_{23},...,x_{2,D}), \\
&&... \\
\overrightarrow{v_{D+2}} &=&(x_{D+2,1},x_{D+2,2},x_{D+2,3},...,x_{D+2,D}).
\end{eqnarray*}

Applying Eq. (\ref{eq2.20}) to $\{\langle \overrightarrow{v_{1}}|%
\overrightarrow{v_{j}}\rangle \}_{j=2}^{D+2},$ we see that $%
\{x_{21},x_{31},...,x_{D+2,1}\}$ are all real and are all in $(-1,0].$ Then $%
\{x_{j1}x_{k1}\}_{2\leq j<k\leq D+2}$ are all real and are all in $[0,1).$
Denote
\begin{eqnarray*}
X_{2} &=&(x_{22},x_{23},x_{24},...,x_{2,D}),..., \\
X_{D+2} &=&(x_{D+2,2},x_{D+2,3},x_{D+2,4},...,x_{D+2,D}), \\
X_{2}^{\ast } &=&(x_{22}^{\ast },x_{23}^{\ast },x_{24}^{\ast
},...,x_{2,D}^{\ast }),..., \\
X_{2}^{\ast }\cdot X_{3} &=&x_{22}^{\ast }x_{32}+x_{23}^{\ast
}x_{33}+...+x_{2D}^{\ast }x_{3D},....
\end{eqnarray*}%
Employing Eq. (\ref{eq2.20}) to $\{\langle \overrightarrow{v_{j}}|%
\overrightarrow{v_{k}}\rangle \}_{2\leq j<k\leq D+2}$, we see that $%
\{X_{j}^{\ast }\cdot X_{k}\}_{2\leq j<k\leq D+2}$ are all real and
\begin{equation}
-1-x_{j1}x_{k1}<X_{j}^{\ast }\cdot X_{k}\leq -x_{j1}x_{k1}.  \label{eq2.22}
\end{equation}%
$\{X_{j}\}_{2\leq j\leq D+2}$ are all nonzero vectors, otherwise if $X_{j}=0$
then $|x_{j1}|=1=|\langle \overrightarrow{v_{1}}|\overrightarrow{v_{j}}%
\rangle |$ contradicting Eq. (\ref{eq2.20}).

We now prove that $\{\frac{X_{j}}{||X_{j}||}\}_{2\leq j\leq D+2}$ satisfies $%
-1<\frac{X_{j}^{\ast }}{||X_{j}||}\cdot \frac{X_{k}}{||X_{k}||}.$ Consider
the contrapositive, without loss of generality, we suppose $\frac{%
X_{2}^{\ast }}{||X_{2}||}\cdot \frac{X_{3}}{||X_{3}||}=-1,$ then
\begin{equation}
X_{3}=-\frac{||X_{3}||}{||X_{2}||}X_{2}.  \label{eq2.23}
\end{equation}
Eq. (\ref{eq2.20}) for $\langle \overrightarrow{v_{2}}|\overrightarrow{v_{3}}%
\rangle ,$ $\langle \overrightarrow{v_{2}}|\overrightarrow{v_{j}}\rangle ,$
and $\langle \overrightarrow{v_{3}}|\overrightarrow{v_{j}}\rangle $ with $%
j>3 $ respectively yields
\begin{eqnarray}
-1 &<&x_{21}x_{31}-||X_{2}||||X_{3}||\leq 0,  \label{eq2.24} \\
-1 &<&x_{21}x_{j1}+X_{2}^{\ast } \cdot X_{j}\leq 0,  \label{eq2.25} \\
-1 &<&x_{31}x_{j1}-\frac{||X_{3}||}{||X_{2}||}X_{2}^{\ast } \cdot X_{j}\leq 0.
\label{eq2.26}
\end{eqnarray}
Since $\{x_{j1}x_{k1}\}_{2\leq j<k\leq D+2}$ are all in $[0,1),$ then Eqs. (%
\ref{eq2.25},\ref{eq2.26}) result in
\begin{equation}
X_{2}^{\ast } \cdot X_{j}=0,x_{21}x_{j1}=x_{31}x_{j1}=0.  \label{eq2.27}
\end{equation}
Combining Eq. (\ref{eq2.23}), we also get
\begin{equation}
X_{3}^{\ast } \cdot X_{j}=0.  \label{eq2.28}
\end{equation}
Thus $x_{21}=x_{31}=0$ or $x_{j1}=0$ for all $j>3.$ If $x_{21}=x_{31}=0$,
Eq. (\ref{eq2.23}) implies that $\overrightarrow{v_{2}}=-\overrightarrow{%
v_{3}},$ $\overrightarrow{v_{2}}$ and $\overrightarrow{v_{3}}$ are all
orthogonal to any vector of $\{\overrightarrow{v_{j}}\}_{j=1}^{D+2}%
\backslash \{\overrightarrow{v_{2}},\overrightarrow{v_{3}}\},$ that is, $\{%
\overrightarrow{v_{j}}\}_{j=1}^{D+2}$ can be further decomposed as $\{%
\overrightarrow{v_{j}}\}_{j=1}^{D+2}=\{\overrightarrow{v_{2}},%
\overrightarrow{v_{3}}\}\cup (\{\overrightarrow{v_{j}}\}_{j=1}^{D+2}%
\backslash \{\overrightarrow{v_{2}},\overrightarrow{v_{3}}\}),$ this
contradicts the hypothesis that $S_{c}$ can not be further decomposed.
Similarly, if $x_{21}x_{31}\neq 0,$ then Eqs. (\ref{eq2.27},\ref{eq2.28})
imply that $x_{j1}=0$ for all $j>3$ and any vector in $\{\overrightarrow{%
v_{j}}\}_{j=1}^{D+2}\backslash \{\overrightarrow{v_{1}},\overrightarrow{v_{2}%
},\overrightarrow{v_{3}}\}$ is orthogonal to any vector in $\{%
\overrightarrow{v_{1}},\overrightarrow{v_{2}},\overrightarrow{v_{3}}\},$
that is, $\{\overrightarrow{v_{j}}\}_{j=1}^{D+2}$ can be further decomposed
as $\{\overrightarrow{v_{j}}\}_{j=1}^{D+2}=\{\overrightarrow{v_{1}},%
\overrightarrow{v_{2}},\overrightarrow{v_{3}}\}\cup (\{\overrightarrow{v_{j}}%
\}_{j=1}^{D+2}\backslash \{\overrightarrow{v_{1}},\overrightarrow{v_{2}},%
\overrightarrow{v_{3}}\}),$ this also contradicts the hypothesis. As a
result, $\{\frac{X_{j}}{||X_{j}||}\}_{2\leq j\leq D+2}$ satisfies $-1<\frac{%
X_{j}^{\ast }}{||X_{j}||}\cdot \frac{X_{k}}{||X_{k}||}.$ Together with Eq. (%
\ref{eq2.22}), hence $\{\frac{X_{j}}{||X_{j}||}\}_{2\leq j\leq D+2}$
satisfies
\begin{equation}
-1<\frac{X_{j}^{\ast }}{||X_{j}||}\cdot \frac{X_{k}}{||X_{k}||}\leq 0.
\label{eq2+1}
\end{equation}

We divide $\{\frac{X_{j}}{||X_{j}||}\}_{2\leq j\leq D+2}$ into a union of
nonempty sets such that any two vectors in distinct sets are orthogonal,
\begin{equation}
\{\frac{X_{j}}{||X_{j}||}\}_{2\leq j\leq D+2}=\cup _{\alpha =1}^{\widetilde{m%
}}T_{\alpha },  \label{eq2+2}
\end{equation}
and each $T_{\alpha }$ can not be decomposed further, as we have done for $\{\overrightarrow{v_{j}}\}_{j=1}^{n}$ in Eq. (\ref%
{eq2.19}). In Eq. (\ref{eq2+2}), each of $\{T_{\alpha }\}_{\alpha =1}^{%
\widetilde{m}}$ has structure (A.a) or (A.c), but not (A.b), since in Eq. (%
\ref{eq2+1}), $-1<\frac{X_{j}^{\ast }}{||X_{j}||}\cdot \frac{X_{k}}{||X_{k}||%
}$ but not $-1\leq \frac{X_{j}^{\ast }}{||X_{j}||}\cdot \frac{X_{k}}{%
||X_{k}||}.$ If $\widetilde{m}\geq 2,$ since $\{x_{j1}x_{k1}\}_{2\leq
j<k\leq D+2}$ are all real and are all in $[0,1),$  with Eq. (\ref{eq2.20}), then there is only one $%
T_{\alpha }$ (without loss of generality, suppose $T_{\alpha }=T_{1})$ in $%
\{T_{\alpha }\}_{\alpha =1}^{\widetilde{m}}$ such that $x_{j1}=0$ if $\frac{%
X_{j}}{||X_{j}||}\notin T_{1}.$ Consequently, $\{\overrightarrow{v_{j}}\}_{j=1}^{D+2}$ can be further
decomposed by $\{\overrightarrow{v_{j}} \Big|\frac{%
X_{j}}{||X_{j}||}\notin T_{1}\}_{2\leq j\leq
D+2},$ this contradicts the hypothesis. Then $\{\frac{X_{j}}{||X_{j}||}%
\}_{2\leq j\leq D+2}$ is of structure (A.c) and can not be decomposed
further. Since each of $\{\frac{X_{j}}{||X_{j}||}\}_{2\leq j\leq D+2}$ has $D-1$
components by definition, then
\begin{equation}
\dim (\text{span}\{\{\frac{X_{j}}{||X_{j}||}\}_{2\leq j\leq D+2}\})\leq D-1.
\label{eq2+3}
\end{equation}%
Eq. (\ref{eq2+3}) contradicts the induction hypothesis that Eq. (\ref{eq2.21}%
) is true for all space dimensions $\dim ($span$\{S_{c}\})=1,2,3,..,D-1$,
with $D\geq 3.$ In conclusion, there do not exist $D+2$ normalized vectors $%
\{\overrightarrow{v_{j}}\}_{j=1}^{D+2}$ satisfying Eq. (\ref{eq2.20}). Then
Eq. (\ref{eq2.21}) holds.

We remark that Lemma 1 can be viewed as a generalization of Lemma 14 in Ref. \cite{Bievre-2022-JMP}.

Notice that, we can also express Lemma 1 in a slightly different form.
When $\{\overrightarrow{w_{j}}\}_{j=1}^{n}\subseteq \mathbb{C}^{d}\backslash \{0\}$
are not necessarily normalized, we can use $\{\frac{\overrightarrow{w}}{||%
\overrightarrow{w_{j}}||}\}_{j=1}^{n}\subseteq \mathbb{C}^{d}\backslash \{0\}$ to
replace $\{\overrightarrow{v_{j}}\}_{j=1}^{n}\subseteq \mathbb{C}^{d}\backslash
\{0\}. $

%


\end{document}